\documentclass[5p, times]{elsarticle}
\usepackage{hyperref}

\journal{Information Sciences}

\usepackage{balance}
\usepackage{listings}
\usepackage{color}

\definecolor{dkgreen}{rgb}{0,0.6,0}
\definecolor{gray}{rgb}{0.5,0.5,0.5}
\definecolor{mauve}{rgb}{0.58,0,0.82}

\makeatletter
\def\lst@makecaption{%
  \def\@captype{table}%
  \@makecaption
}
\makeatother

\usepackage{tikz}
\usepackage{caption}
\usepackage{pifont}
\DeclareFixedFont{\ttb}{T1}{txtt}{bx}{n}{9} 
\DeclareFixedFont{\ttm}{T1}{txtt}{m}{n}{9}  

\usepackage{color}
\definecolor{deepblue}{rgb}{0,0,0.5}
\definecolor{deepred}{rgb}{0.6,0,0}
\definecolor{deepgreen}{rgb}{0,0.5,0}

\usepackage{listings}

\newcommand\pythonstyle{\lstset{
language=Python,
basicstyle=\small,
otherkeywords={self, virtualenv, pip, source},             
keywordstyle=\ttb\color{deepblue},
emph={MyClass,__init__, python3, test},          
emphstyle=\ttb\color{deepred},    
stringstyle=\ttb\color{deepgreen},
frame=tb,                         
showstringspaces=false            %
}}

\lstnewenvironment{python}[1][]
{
\pythonstyle
\lstset{#1}
}
{}

\newcommand\pythoninline[1]{{\pythonstyle\lstinline!#1!}}

\usepackage{tikz}

\usepackage{amssymb}
\usepackage{pifont}

\graphicspath{{img/}}

\usepackage{url}
\usepackage{multirow}

\usepackage{amssymb}
\usepackage{amsmath}

\usepackage{color-edits}

\addauthor{pg}{red}
\addauthor{dh}{orange}
\addauthor{ef}{blue}

\usepackage{algorithmic}
\usepackage{array}

\usepackage[caption=false,font=footnotesize,labelfont=sf,textfont=sf]{subfig}

\begin{document}
\begin{frontmatter}

\title{Benchmarks for Graph Embedding Evaluation}


\author{Palash Goyal\corref{co}}
\author{Di Huang\corref{co}}
\author{Ankita Goswami\corref{}}
\author{Sujit Rokka Chhetri\corref{}}
\author{Arquimedes Canedo\corref{}}
\author{Emilio Ferrara\corref{}}
\cortext[co]{These authors contributed equally to this work.}

\address{University of Southern California, Information Sciences Institute\\4676 Admiralty Way, Suite 1001. Marina del Rey, CA. 90292, USA}

\begin{abstract}

Graph embedding is the task of representing nodes of a graph in a low-dimensional space and its applications for graph tasks have gained significant traction in academia and industry. Variability in the embedding methods is often due to the different properties preserved by them. Majority of methods report performance boosts on few selected real graphs and the results are thus not generalizable. In this work, we introduce a principled framework to compare them. Our goal is threefold: (i) provide a framework to compare the performance of various graph embedding methods, (ii) establish a benchmark with real-world graphs that exhibit different structural properties, and (iii) provide users with a tool to identify the best embedding method for their data. We evaluate the most influential graph embedding methods and traditional link prediction methods against a corpus of 100 real-world networks with varying properties. We organize the networks based on their properties to draw insights into embedding performance. We define GFS-score to quantify the performance of an approach and use it to rank the state-of-the-art embedding approaches. We envision that the proposed framework (https://www.github.com/palash1992/GEM-Benchmark) will serve the community as a benchmarking platform to test and compare the performance of future graph embedding techniques.

\end{abstract}


\end{frontmatter}


\section{Introduction}
Graphs are a natural way to represent relationships and interactions between entities in real systems. For example, people on social networks, proteins in biological networks, and authors in publication networks can be represented by nodes in a graph, and their relationships such as friendships, protein-protein interactions, and co-authorship are represented by edges in a graph. These graphical models enable us to understand the behavior of  systems and to gain insight into their structure. These insights can further be used to predict future interactions and missing information in the system. These tasks are formally defined as link prediction and node classification. Link prediction estimates the likelihood of a relationship among two entities. This is used, for example, to recommend friends on social networks and to sort probable protein-protein interactions on biological networks. Similarly, node classification estimates the likelihood of a node's label. This is used, for example, to infer missing meta-data on social media profiles, and genes in proteins.

Numerous graph analysis methods have been developed. Broadly, these methods can be categorized as non-parametric and parametric. Non-parametric methods operate directly on the graph whereas parametric methods represent the properties of nodes and edges in the graph in a low-dimensional space. Non-parametric methods such as Common Neighbors~\cite{newman2001clustering}, Adamic-Adar~\cite{adamic2003friends} and Jaccard's coefficient~\cite{salton1986introduction} require access and knowledge of the entire graph for the prediction. On the other hand, parametric models such as Thor et. al.~\cite{thor2011link} employ graph summarization and define super nodes and super edges to perform link prediction. Kim et. al.~\cite{kim2011network} use Expectation Maximization to fit the real network as a Kronecker graph and estimate the parameters. Another class of parametric models that have gained much attention recently are graph embeddings~\cite{GOYAL2018,hamilton2017representation,cai2018comprehensive}. Graph embedding methods define a low-dimensional vector for each node and a distance metric on the vectors. These methods learn the representation by preserving certain properties of the graph. Graph Factorization~\cite{Ahmed2013} preserves visible links, HOPE~\cite{Wang2016} aims to preserve higher order proximity, and node2vec~\cite{Grover2016} preserves both structural equivalence and higher order proximity. In this paper, we focus our attention on graph embedding methods. While this is a very active area of research that continues to gain popularity among researchers, there are several challenges that must be addressed before graph embedding algorithms become mainstream.

\subsection{Challenges}\label{sec:challenges}
Most research on graph embedding has focused on the development of mechanisms to preserve various characteristics of the graph in the low-dimensional space. However, very little attention has been dedicated to the development of mechanisms to rigorously compare and evaluate different graph embedding methods. To make matters worse, most of the existing work use simple synthetic data sets for visualization and a few real networks for quantitative comparison. Goyal et. al.~\cite{GOYAL2018} use Stochastic Block Models to visualize the results of graph embedding methods. Salehi at. al.~\cite{salehi2017properties} use the Barabasi-Albert graph to understand the properties of embeddings. Such evaluation strategy suffers from the following challenges:
\begin{enumerate}
    \item Properties of real networks vary according to the domain. Therefore it is often difficult to ascertain the reason behind the performance improvement of a given method on a particular real dataset (as shown in ~\cite{GOYAL2018}).
    \item As demonstrated in this paper, the performance of embedding approaches vary greatly, and according to the properties of different graphs. Therefore, the utility of any specific method is difficult to establish and to characterize. In practice, the performance improvement of a method can be attributed to stochasticity.
    \item Different methods use different metrics for evaluation. This makes it very difficult to compare the performance of different graph embedding methods on a given problem.
    \item Typically, each graph embedding method has a reference implementation. This implementation makes specific assumptions about the data, representation, etc. This further complicates the comparison between methods.
    
\end{enumerate}

\subsection{Contributions}

In this work, we aim to: (i) provide a unifying framework for comparing the performance of state-of-the-art and future graph embedding methods; (ii) establish a benchmark comprised of 100 real-world graphs that exhibit different structural properties; and (iii) provide users with a fully automated Python library that selects the best graph embedding method for their graph data. We address the above challenges (Section~\ref{sec:challenges}) with the following contributions:

\begin{enumerate}
    \item \sloppypar{We propose an evaluation benchmark to compare and evaluate embedding methods. \textcolor{black}{This benchmark consists of 100 real-world graphs, largely a subset of the CommunityFitNet corpus~\cite{ghasemian:etal:2019} with a few additional networks drawn from the Stanford Network Analysis Project (SNAP)~\cite{leskovec2016snap}. The benchmark categorizes networks in four domains: social, biology, technological and economic, based on the taxonomy provided by the Index of Complex Networks (ICON)~\cite{clauset:etal:2016}.}}
    \item Using our evaluation benchmark, we evaluate and compare 8 state-of-the-art methods and provide, for the first time, a characterization of their performance against graphs with different properties. We also compare their scores with traditional link prediction methods and ascertain the general utility of embedding methods.
    \item A new score, GFS-score, is introduced to compare various graph embedding methods for link prediction. The GFS-score provides a robust metric to evaluate a graph embedding approach by averaging over 100 graphs. It further has many components based on the type and property of graph yielding insights into the methods.
    \item A Python library comprised of 4 state-of-the-art embedding methods, and 4 traditional link prediction methods. This library automates the evaluation, comparison against all the other methods, and performance plotting of any new graph embedding method.
\end{enumerate}

\subsection{Organization}
The rest of the work is organized as follows. Section~\ref{sec:notation} presents the notations used in the paper and and overview of graph embeddings. Section~\ref{sec:gem-ben} introduces the benchmark framework, describes the real benchmark graphs, defines the GFS-score. Section~\ref{sec:results} presents the results and analysis. Section~\ref{sec:library} introduces the Python library. Section~\ref{sec:conclusion} concludes.

\section{Notations and Background}~\label{sec:notation}
This section introduces the notation used in this paper, and provides a brief overview of graph embedding methods. For an in-depth analysis of graph embedding theory we refer the reader to \cite{GOYAL2018}.

\subsection{Notations}
\begin{table}[!htbp]
    \footnotesize
    \centering
    \renewcommand{\arraystretch}{1.3}
    \caption{Summary of notation}
    \label{tab:not}
        \begin{tabular}{c|c}
            \hline \hline
             $G$ & Graphical representation of the data \\ \hline
             $V$ & Set of vertices in the graph \\ \hline
             $E$ & Set of edges in the graph \\ \hline
             $W$ & Adjacency matrix of the graph, $|V| \times |V|$ \\ \hline
             $f$ & Embedding function \\ \hline
             $\mathcal{S}$ & Set of synthetic graphs  \\ \hline
             $\mathcal{R}_D$ & Set of real graphs in domain $D$  \\ \hline
             $\mathcal{D}$ & Set of domains  \\ \hline
             $\mathcal{M}$ & Set of evaluation metrics  \\ \hline
             $e_m$ & Evaluation function for metric $m$ \\ \hline
             $\mathcal{A}$ & Set of graph and embedding attributes \\ \hline
             $d$ & Number of embedding dimensions \\ \hline
             $Y$ & Embedding of the graph, $|V| \times d$ \\  \hline \hline
        \end{tabular}
\end{table}

$G (V, E)$ denotes a weighted graph where $V$ is the set of vertices and $E$ is the set of edges. We represent $W$ as the adjacency matrix of $G$, where $W_{ij} = 1$ represents the presence of an edge between $i$ and $j$. A graph embedding is a mapping $f: V -> \mathbb{R}^d$, where $d  << |V|$ and the function $f$ preserves some proximity measure defined on graph $G$. It aims to map similar nodes close to each other. Function $f$ when applied on the graph $G$ yields an embedding $Y$.

In this work, we evaluate four state-of-the-art graph embedding methods on a set of real graphs denoted by $\mathcal{R}$ and synthetic graphs denoted by $\mathcal{S}$ (the results for synthetic graphs are presented in the Appendix). To analyze the performance of methods, we categorize the graphs into a set of domains $\mathcal{D} = \lbrace$ Social, Economic, Biology, Technological$\rbrace$. The set of graphs in a domain $D \in \mathcal{D}$ is represented as $\mathcal{R}_D$. We use multiple evaluation metrics on graph embedding methods to draw insights into each approach. We denote this set of metrics as $\mathcal{M}$. The notations are summarized in Table~\ref{tab:not}.

\subsection{Graph Embedding Methods}
Graph embedding methods embed graph vertices into a low-dimensional space. The goal of graph embedding is to preserve certain properties of the original graph such as distance between nodes and neighborhood structure. Based upon the function $f$ used for embedding the graph, existing methods can be classified into three categories~\cite{GOYAL2018}: factorization based, random walk based and deep learning based. 

\subsubsection{Factorization based approaches}
Factorization based approaches apply factorization on graph related matrices to obtain the node representation. Graph matrices such as the adjacency matrix, Laplacian matrix, and Katz similarity matrix contain information about node connectivity and the graph's structure. Other matrix factorization approaches use the eigenvectors from spectral decomposition of a graph matrix as node embeddings. For example, to preserve locality, LLE~\cite{Roweis2000} uses $d$ eigenvectors corresponding to eigenvalues from second smallest to $(d+1)^{th}$ smallest from the sparse matrix $(I-W)^\intercal(I-W)$. It assumes that the embedding of each node is a linear weighted combination of the neighbor's embeddings. Laplacian Eigenmaps ~\cite{belkin2001laplacian} take the first $d$ eigenvectors with the smallest eigenvalues of the normalized Laplacian $D^{-1/2}LD^{-1/2}$. Both LLE and Laplacian Eigenmaps were designed to preserve the local geometric relationships of the data. Another type of matrix factorization methods learn node embeddings under different optimization functions in order to preserve certain properties. Structural Preserving Embedding~\cite{shaw2009structure} builds upon Laplacian Eigenmaps to recover the original graph. Cauchy Graph Embedding~\cite{luo2011cauchy} uses a quadratic distance formula in the objective function to emphasize similar nodes instead of dissimilar nodes. Graph Factorization~\cite{Ahmed2013} uses an approximation function to factorize the adjacency matrix in a more scalable manner. GraRep~\cite{Cao} and HOPE~\cite{Ou2016} were invented to keep the high order proximity in the graph. Factorization based approaches have been widely used in practical applications due to their scalability. The methods are also easy to implement and can yield quick insights into the data set.

\subsubsection{Random walk approaches}
\sloppypar{Random walk based algorithms are more flexible than factorization methods to explore the local neighborhood of a node for high-order proximity preservation. DeepWalk~\cite{Perozzi2014} and Node2vec~\cite{Grover2016} aim to learn a low-dimensional feature representation for nodes through a stream of random walks. These random walks explore the nodes' variant neighborhoods. Thus, random walk based methods are much more scalable for large graphs and they generate informative embeddings. Although very similar in nature, DeepWalk simulates uniform random walks and Node2vec employs search-biased random walks, which enables embedding to capture the community or structural equivalence via different bias settings. LINE~\cite{Tang2015} combines two phases for embedding feature learning: one phase uses a breadth-first search (BFS) traversal across first-order neighbors, and the second phase focuses on sampling nodes from second-order neighbors. HARP~\cite{chen2017harp} improves DeepWalk and Node2vec by creating a hierarchy for nodes and using the embedding of the coarsened graph as a better initialization in the original graph. Walklets ~\cite{perozzi2016walklets} extended Deepwalk by using multiple skip lengths in random walking. Random walk based approaches tend to be more computationally expensive than factorization based approaches but can capture complex properties and longer dependencies between nodes.}

\subsubsection{Neural network approaches}
The third category of graph embedding approaches is based on neural networks. Deep neural networks based approaches capture highly non-linear network structure in graphs, which is neglected by factorization based and random walk based methods. One type of deep learning based methods such as \textit{SDNE}~\cite{Wang2016} uses a deep autoencoder to provide non-linear functions to preserve the first and second order proximities jointly. Similarly, \textit{DNGR}~\cite{cao2016deep} applies random surfing on input graph before a stacked denoising autoencoder and makes the embedding robust to noise in graphs. Another genre of methods use Graph Neural Networks(\textit{GNNs}) and  Graph Convolutional Networks (\textit{GCNs})~\cite{bruna2013spectral, henaff2015deep, li2015gated, hamilton2017inductive} to aggregate the neighbors embeddings and features via convolutional operators, including spatial or spectral filters. GCNs learn embeddings in a semi-supervised manner and have shown great improvement and scalability on large graphs compared to other methods. \textit{SEAL}~\cite{zhang2018link} learns a wide range of link prediction heuristics from extracted local enclosing subgraphs with GNN. \textit{DIFFPOOL}~\cite{ying2018hierarchical} employs a differentiable graph pooling module on GNNs to learn hierarchical embeddings of graphs. Variational Graph Auto-Encoders(\textit{VGAE})~\cite{kipf2016variational} utilizes a GCN as encoder and inner product as decoder, which provides embedding with higher quality than autoencoders. Deep neural network based algorithms like \textit{SDNE} and \textit{DNGR} can be computational costly since they require the global information such as adjacency matrix for each node as input. GCNs based methods are more scalable and flexible to characterize global and local neighbours through variant convolutional and pooling layers.

\section{GEM-BEN: Graph Embedding Methods Benchmark}~\label{sec:gem-ben}
 \begin{figure*}[htpb]
   \centering
  \includegraphics[width=0.95\textwidth]{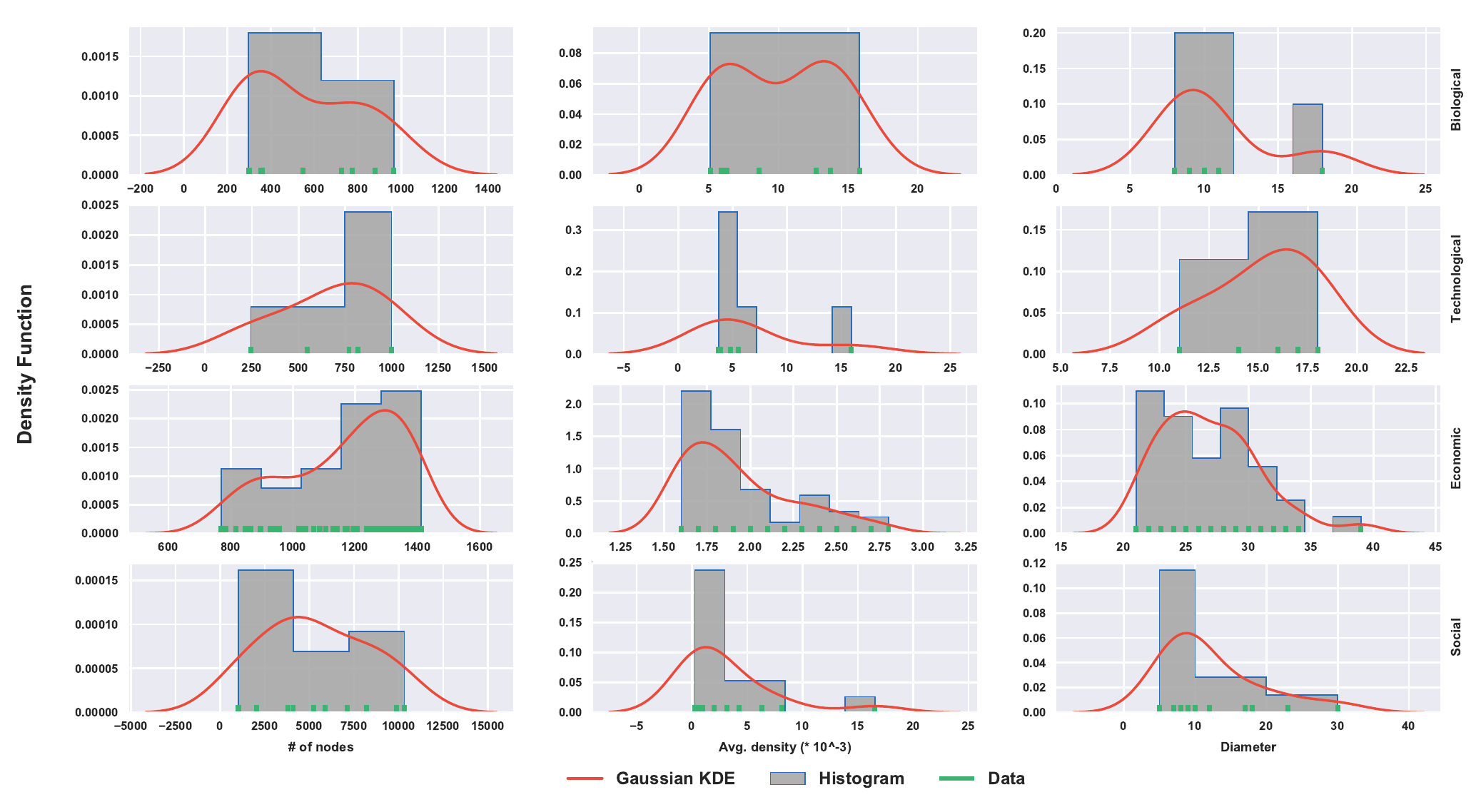}
  \caption{Real graphs properties.}
  \label{fig:realProp}
\end{figure*}
Unlike other fields with well established benchmark datasets (e.g. community detection~\cite{lancichinetti2008benchmark}), the graph embedding community has adopted an ad-hoc approach to evaluate new methods. Typically, graph embedding methods are evaluated on only a few real networks, and these are biased towards specific properties. This ad-hoc evaluation approach restricts us from understanding how the algorithm would behave if we vary a certain property of the graph, or how the algorithm performs on other types of graphs. In order to propose a more rigorous evaluation approach, we must first to understand the key attributes that govern the performance of graph embedding methods. First, the \textit{size of the graph} (\textbf{A1}) is a challenge for any method. Real graphs vary in the number of nodes, from a few hundred to millions of nodes. Different methods make different assumptions on how to capture the higher order proximities and structural dependencies between nodes, and this greatly affects their scalability. Second, the \textit{density of the graph} (\textbf{A2}) plays an important role in defining its structure. Lower density results in lesser information about the nodes which may hamper the performance of some methods. Third, the \textit{dimension of the embedding} (\textbf{A3}) determines how concisely the method can store the information about a given graph. Higher dimension of the embedding may lead to overfitting of the graph whereas lower dimension of the embedding may not be enough to capture the information the graph provides resulting in underfitting. Fourth, the \textit{evaluation metric} (\textbf{A4}) used to evaluate the method captures different aspects of the prediction. Global metrics are often biased towards high degree nodes whereas local metrics can be biased towards lower degree nodes.


In this paper, we take the first step towards establishing a graph embedding benchmark. We propose a benchmark evaluation framework to answer the following questions: 
\begin{itemize}
    \item Q1: How does the performance of embedding methods vary with the increasing size of the graph?
    \item Q2: How does increasing the density of graph affect the model?
    \item Q3: How does the optimal embedding dimension vary with an increasing number of nodes in the graph?
    \item Q4: How does the performance vary with respect to the evaluation metric?
\end{itemize}

\textcolor{black}{To address the above questions, we construct and utilize a benchmark of 100 real-world graphs, drawn from two large existing network corpora~\cite{ghasemian:etal:2019,leskovec2016snap}, and vary the above attributes (A1, ..., A4) in the graphs and the embedding methods.}
Varying the \textit{size of the graph} (\textbf{A1}) in terms of number of nodes answers the first question (Q1) and helps us understand which methods are best when used in small, medium, and large graphs. Similarly, varying the \textit{density of the graph} (\textbf{A2}) in terms of the average degree of nodes helps us understand its effect in the embedding performance. This answers the second question (Q2). Furthermore, varying the \textit{dimension of the embedding} (\textbf{A3}) helps us draw insights into the information compression power of the embedding approach. This answers the third question (Q3). Finally, by varying the \textit{evaluation metrics} (\textbf{A4}) we can analyze the performance sensitivity of the method and can help us infer the bias of the embedding method towards specific nodes int he graph. This answers the fourth question (Q4).


\subsection{Real Graphs}
\textcolor{black}{We use a benchmark corpus that contains 100 real-world graphs, 86 of which are from the CommunityFitNet corpus~\cite{ghasemian:etal:2019} and 14 are from the Stanford Network Analysis Project (SNAP)~\cite{leskovec2016snap}. These graphs are labeled by their domain of origin as social, biological, economic, and technological, following the taxonomy given by the Index of Complex Networks (ICON)~\cite{clauset:etal:2016}.} We use Induced Subgraph Random Walk Sampling (ISRW)~\cite{lu2012sampling} to sample some huge social graphs while keeping the same graph density to ensure.  To demonstrate the usefulness of this benchmark, we evaluate eight graph embedding methods and measure their performance. This provides valuable insights about every method and their sensitivity to different graph properties. This paves the way towards a framework that can be used to recommend the best embedding approaches for a given graph with a unique set of properties.


Figure~\ref{fig:realProp} summarizes the main properties of the graphs from different domains in the data set. We observe that economic graphs have a lower average density varying between 0.00160 and 0.00280 with a higher number of nodes concentrated in lower density spectrum. Technological and social graphs are denser with an average density between 0.0030 to 0.0160. It is interesting to note that despite the wide average density range densities are concentrated primarily in the lower and higher values with a gap in between. Biological graphs have an almost uniform distribution of densities ranging from 0.005 to 0.0155.

Next, we observe the domain-wise pattern of diameters. Economic graphs have the widest range(20 - 40) and the highest values of diameters which justifies the lowest average densities observed. Technological graphs with diameter ranges between 11 and 17.5 are less sparse when compared with economic graphs. Biological graphs have a good combination of both dense and sparse graphs with a majority of graphs lying in small diameter range. Biological graphs typically have short long diameter ranges as (8 to 12) and (16 to 18) respectively. Social graphs have in general a lower diameter around 10 although some of them have higher diameters.

On further investigation, we observe that biological networks have the highest clustering tendencies with an average clustering coefficient as 0.10. However, economic graphs stand in absolute contrast to them with very low clustering coefficient of 0.00016 as the highest recorded average clustering coefficient. Technological networks are somewhere in between the aforementioned extremes with 0.03 as the highest recorded average clustering coefficients. Clustering tendencies can be sought to have a high correlation with average density and diameter observations.

 Note that these 100 graphs include a very diverse set of graphs in terms of the \textit{size of the graph} (\textbf{A1}) ranging from 200 to 1500 nodes, and in terms of the \textit{density of the graph} (\textbf{A2}) ranging from an average density between 0.0015 to 0.020. As it will be shown in Section~\ref{sec:results}, this graph diversity is helpful in characterizing the performance of different embedding methods.












\subsection{Evaluation Metrics}
\sloppypar{
In the graph embedding literature, there are two primary metrics that are used to evaluate the performance of the methods on link prediction: (i) Precision at $k$ ($P@k$) and (ii) Mean Average Precision ($MAP$).} These metrics are defined as follows:

\textbf{\textit{P@k}} is the fraction of correct predictions in the top $k$ predictions. It is defined as 
          $P@k = \frac{|E_{pred}(1:k) \cap E_{obs}|}{k}$,
        where $E_{pred}(1:k)$ are the top $k$ predictions and $E_{obs}$ are the observed edges/hidden edges. 

\textbf{\textit{MAP}} estimates the prediction precision for every node and computes the prediction average over all nodes, as follows:
        \begin{equation*}
            MAP = \frac{\sum_i AP(i)}{|V|},
        \end{equation*}
        where $AP(i) = \frac{\sum_k P@k(i) \cdot \mathbb{I}\{E_{pred_i}(k) \in E_{obs_i}\}}{|\{k: E_{pred_i}(k) \in E_{obs_i}\}|}$, $P@k(i) = \frac{|E_{pred_i}(1:k) \cap E_{obs_i}|}{k}$, 
        and $E_{pred_i}$ and $E_{obs_i}$ are the predicted and observed edges for node $i$ respectively.
        
Intuitively, $P@k$ is a global metric that measures the accuracy of the most likely links predicted. On the other hand, \textit{MAP} measures the accuracy of prediction for each node and computes their average. These metrics are often uncorrelated and reflect the properties captured by the prediction method at different levels ($MAP$ on local level and $P@k$ on global level). In this work, we present results using both these metrics to analyze each approach.

\subsection{GFS-score}
We now define a set of scores to evaluate a graph embedding model on our data set. The scores are divided into components to draw insights into a method's approach across domains and metrics. We further plot the metrics varying various graph properties to understand the sensitivity of the models to these properties.

Given a set of graph domains $\mathcal{D}$, a set of evaluation metrics $\mathcal{M}$ and evaluation function $e_m (graph, approach)$ for $m \in \mathcal{M}$, we define GFS-score for an approach $a$ as follows:
\begin{equation}
    micro-GFS-m(a) = \frac{ \sum_{g \in \mathcal{G}} (e_m(g, a)/e_m(g, random)) }{|\mathcal{G}|},
\end{equation}
\begin{equation}
    macro-GFS-m(a) = \frac{\sum_{d \in \mathcal{D}} GFS-m(d, a)}{|\mathcal{D}|},
\end{equation}
\begin{equation}
    GFS-m(d, a) = \frac{ \sum_{g \in \mathcal{G}_d} (e_m(g, a)/e_m(g, random)) }{|\mathcal{G}_d|},
\end{equation}
where $\mathcal{G}_d$ is the set of graphs in domain $d$.

The GFS-score is a robust score which averages over a set of real graphs with varying properties. It is normalized in order to ascertain the gain in performance with respect to a random prediction. The domain scores provide insights into the applicability of each approach to the different graph categories.



\subsection{Link Prediction Baselines}
Our link prediction baselines were selected to showcase the utility of embedding approaches on real graphs and establish the ground truth for comparison between the state-of-the-art methods. The link prediction baselines are:

\textbf{Preferential Attachment}:~\cite{barabasi1999emergence} is based on the assumption that the connection to a node is proportional to its degree. It defines the similarity between the nodes as the product of their degrees.

\textbf{Common Neighbors}:~\cite{newman2001clustering} defines the similarity between nodes as the number of common neighbors between them.

\textbf{Adamic-Adar}:~\cite{adamic2003friends} is based on the intuition that common neighbors with very large neighbourhoods are less significant than common neighbors with small neighborhoods when predicting a connection between two nodes. Formally, it is defined as the sum of the inverse logarithmic degree centrality of the neighbours shared by the two nodes.

\textbf{Jaccard's Coefficient}:~\cite{jaccard1908nouvelles} measures the probability that two nodes $i$ and $j$ have a connection to node $k$, for a randomly selected node $k$ from the neighbors of $i$ and $j$ .

\subsection{Embedding Approaches}
We illustrate the benchmark data set on four popular graph embedding techniques to illustrate the utility of the benchmark and rank the state-of-the-art embedding approaches. The techniques preserve various properties including local neighborhood, higher order proximity and structure.

\textbf{Laplacian Eigenmaps} \cite{belkin2001laplacian}: It penalizes the weighted square of distance between neighbors. This is equivalent to factorizing the normalized Laplacian matrix.  

\textbf{Graph Factorization} (GF) \cite{Ahmed2013}: It factorizes the adjacency matrix with regularization.

\textbf{Higher Order Proximity Preserving} \cite{Ou2016} (HOPE): It factorizes the higher order similarity matrix between nodes using generalized singular value decomposition.

\textbf{Structural Deep Network Embedding} (SDNE) \cite{Wang2016}: It uses deep autoencoder along with Laplacian Eigenmaps objective to preserve first and second order proximities.

\section{Experiments and Analysis}~\label{sec:results}


\begin{table*}[t]
\centering
\caption{Average and standard deviation of GFS-score}
\label{tab:perf_summ}
\begin{tabular}{c|c|c|c|c|c|c|c|c|c|c|c|c}
\multicolumn{1}{l|}{} & \multicolumn{2}{c|}{\textit{micro-GFS}} & \multicolumn{2}{c|}{\textit{macro-GFS}} & \multicolumn{2}{c|}{\textit{GFS-bio}}& \multicolumn{2}{c|}{\textit{GFS-eco}} & \multicolumn{2}{c|}{\textit{GFS-soc}} & \multicolumn{2}{c}{\textit{GFS-tech}}\\ \hline
& $MAP$ & $P@100$ & $MAP$ & $P@100$ & $MAP$ & $P@100$ & $MAP$ & $P@100$ & $MAP$ & $P@100$ & $MAP$ & $P@100$\\ \hline
Pref. Attach. & 3.1 & 37.7 & 3.4 & 31.3 & 3.3 & 35.9 & 2.6 & 2.7 & 6.4 & 83.6 & 1.3 & 3.0\\ \hline
Common Neigh.  & 2.8 & 77 & 4.6 & 67.2 & 3.2 & 36.1 & 0.1 & 0.0 & 14.8 & 232.8 & 0.5 & 0.0\\ \hline
Jaccard Coeff.  & 2.3 & 28.3 & 3.8 & 26.3 & 2.2 & 0.5 & 0.01 & 0.0 & 12.6 & 102.8 & 0.5 & 1.8\\ \hline
Adamic-Adar   & 2.9 & 74.6 & 4.8 & 66.0 & 3.3 & 28.3 & 0.02 & 0.0 & 15.5 & 234.6 & 0.5 & 1.0\\ \hline
Lap. Eigen. & 6.6 & 18.9 & 6.4 & 17.4 & 3.5 & 0.75 & 6.2 & 0.0 & 12.3 & 65.3 & 3.8 & 3.5\\ \hline
GF & 5.7 & 41.2 & 5.2 & 40.9 & 3.3 & 19.2 & 5.7 & 80.5 & 10.5 & 62.2 & 1.35 & 3.4\\ \hline
HOPE & 6.1 & \textbf{98.0} & 7.3 & 89.1 & 4.7 & \textbf{43.8} & 4.2 & 45.2 & 16.7 & \textbf{263.3} & 3.4 & 4.25\\ \hline
SDNE & \textbf{11.0} & 90.6 & \textbf{10.1} & \textbf{91.3} & \textbf{6.8} & 33.1  & \textbf{11.2} & \textbf{143.3} & \textbf{18.6} & 170.4 & \textbf{4.0} & \textbf{18.5}\\ \hline
\end{tabular}
\end{table*}

 \begin{figure*}
   \centering
  \includegraphics[width=0.98\textwidth]{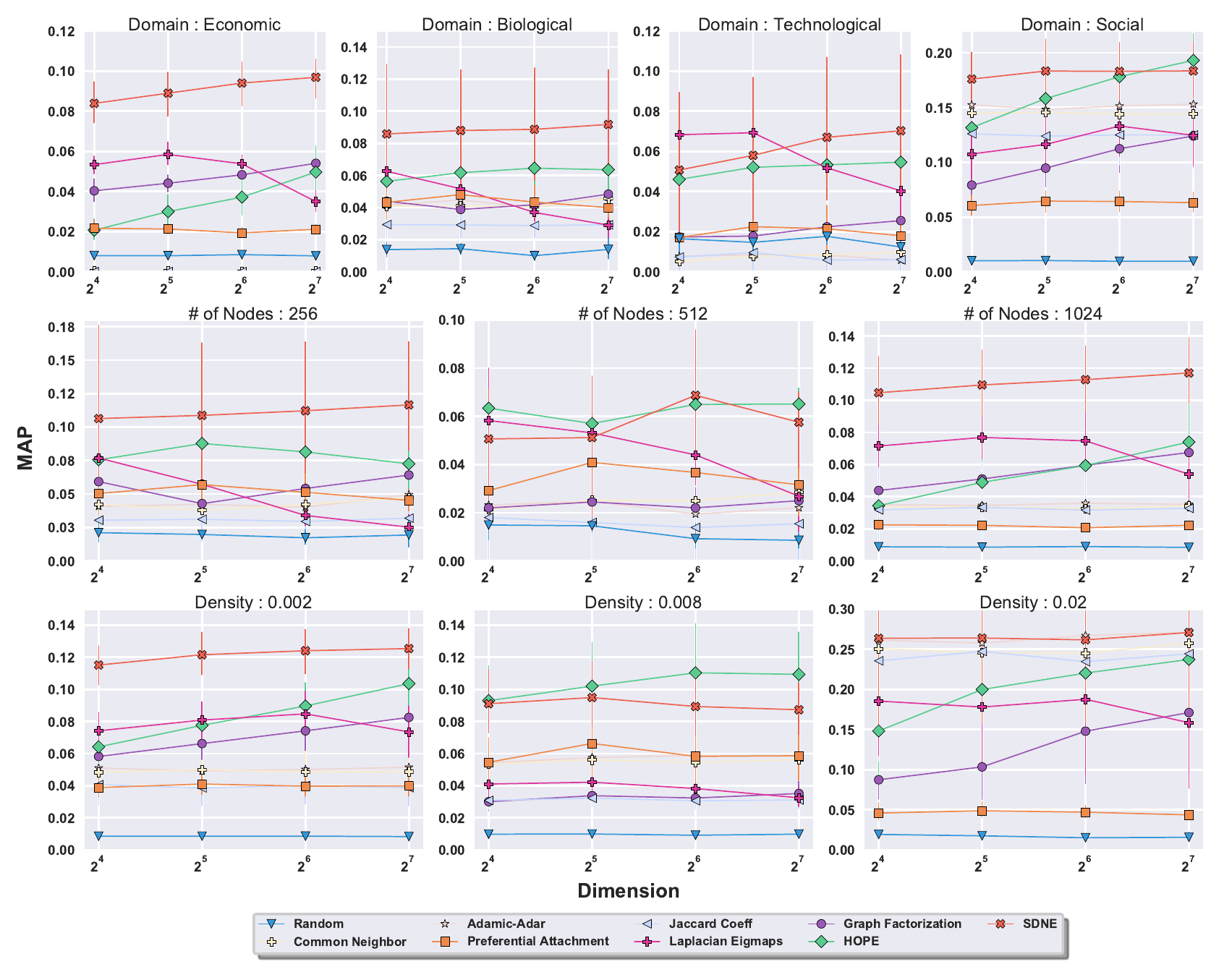}
  \caption{Performance evaluation of different methods varying the attributes of graphs. The x axis denotes the dimension of embedding, whereas the y axis denotes the $MAP$ scores.}
  \label{fig:ben_real}
\end{figure*}

 \begin{figure*}
   \centering
  \includegraphics[width=0.98\textwidth]{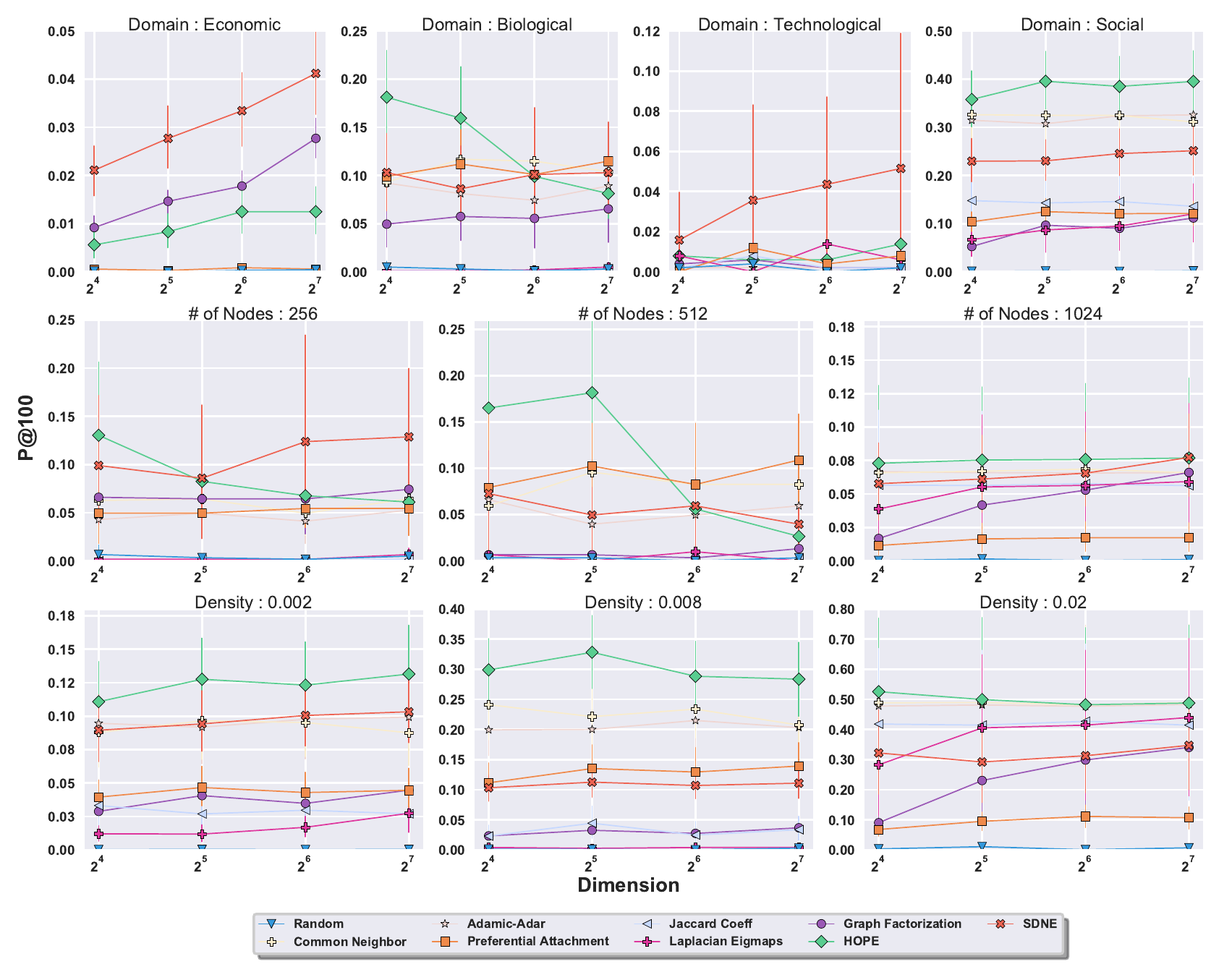}
  \caption{Performance evaluation of different methods varying the attributes of graphs. The x axis denotes the dimension of embedding, whereas the y axis denotes the $P@100$ scores.}
  \label{fig:ben2_real}
\end{figure*}

This section evaluates the performance of the baseline and state-of-the-art methods on link prediction on the benchmark graphs according to the methodology presented in Section~\ref{sec:gem-ben}. First, we present the general results, and use subsections for an in-depth analysis.

Figure~\ref{fig:ben_real} shows the $MAP$ scores achieved by the eight methods when varying the \textit{size of the graph} (\textbf{A1}) (256, 512, and 1024 nodes), the \textit{density of the graph} (\textbf{A2}) (degree 3, 4, and 5), and the \textit{dimension of the embedding} (\textbf{A3}) (dimensions of $2^4, ..., 2^7$). We present the results for four graph categories: economic, biological, social and technological. To ensure the experiments on an attribute are independent of other attributes, while varying the number of nodes, we keep the density within a small range. For density, we consider all the graphs as it is independent of number of nodes. The $MAP$ scores shown are precision scores averaged over all nodes, thus making the score unbiased towards high degree nodes. However, as most real world graphs have a long tail degree distribution it may be unfair to methods which predict the top links in the graphs but fail for nodes with less information. Overall, we observe that methods show consistent performance across data sets from the same domain as shown by the low variance. Further, the best embedding approaches obtain a $MAP$ value of around 0.1 which is about 5 times improvement over the traditional link prediction methods.

Similarly, Figure~\ref{fig:ben2_real} shows the \textit{P@100} values for the eight link prediction methods using the same experimental setup. \textit{P@k} is a global metric which computes the accuracy of top $k$ predictions. As for MAP, the value of \textit{P@100} is consistent across data sets. The best embedding approaches obtain 0.15-0.2 \textit{P@100} which is an improvement of 5-10 times over traditional link prediction methods.

We address the \textit{evaluation metrics} (\textbf{A4}) in Table~\ref{tab:perf_summ} where the GFS-score for each method is presented. In general, SDNE obtains top performance across domains and metrics with the exception of \textit{P@100} in biology and social for which HOPE performs better. Also, we can see that in general embedding approaches outperform traditional methods. We now study the performance variations between methods in detail.

\subsection{Domain Performance}
For economic graphs, we find that traditional link prediction models perform poorly because these graphs do not have the notion of a community structure and therefore violate the assumptions of traditional methods. However, state-of-the-art methods perform better and capture the structure of these networks achieving good scores on both $MAP$ and $P@100$. We observe that Structural Deep Network Embedding (SDNE) gives the best performance for economic graphs with a GFS-score of 11.2 for $MAP$ and 143.3 for $P@100$. The score for $P@100$ is significantly higher than $MAP$. The reason is that predicting correct links for low degree nodes is more challenging than high degree nodes and thus averaging over all nodes gives a low overall performance. We also find that Graph Factorization (GF), despite capturing first order proximity, performs better than Higher Order Proximity Embedding (HOPE). This suggests that economic graphs do not benefit from capturing longer dependency. The overall good performance of SDNE indicates that economic graphs have complex structures that benefit from deep neural networks. 

Biological graphs show a more consistent performance across traditional methods. Preferential Attachment, Common Neighbors and Adamic Adar give a GFS-map score of about 3.3 and a GFS-P@100 score of around 30. Among the graph embedding methods, Graph Factorization and Laplacian Eigenmaps do not improve over the traditional methods and give similar scores. However, HOPE and SDNE show significant gains in performance. SDNE and HOPE give the highest $MAP$ and $P@100$ performance, respectively. As HOPE captures higher order proximity, these results suggest that longer dependencies are critical to predicting top links in the biology domain. For the $MAP$ score, SDNE performs better than HOPE. This may be because the low degree nodes may require the understanding of structure to make accurate predictions. Furthermore, we see that the GFS scores for the biology domain are lower than economy graphs.  This is an indication that these networks are more difficult to predict.

Technological graphs show patterns similar to economic graphs. Non-parametric link prediction methods perform poorly and are unable to outperform the random baseline. Laplacian Eigenmaps gives a score of around 3.5 for both $MAP$ and $P@100$. HOPE and SDNE perform the best among the state-of-the-art methods showing that both higher order and structure are helpful in the prediction. We also observe that for technological graphs, the gain in performance is much higher for $P@100$ by SDNE. This suggests that several top links are not based on neighborhood but based on distant nodes with similar structure.

For social graphs, we observe that traditional methods perform very well mostly because these methods were designed for such graphs. On average, we see that both SDNE and HOPE outperform baselines for $MAP$ and HOPE achieves the highest performance on $P@100$. Social graphs often have community structure and capturing local and global neighborhood can be useful in capturing the graph. Higher performance by HOPE may be because of its ability to capture such properties well.

In general, we observe that the state-of-the-art graph embedding methods have higher micro and macro GFS scores than traditional methods showcasing the utility of these approaches. It is worth noting that Laplacian Eigenmaps is good for predicting links for nodes with lower degrees, but it does not perform well for top link predictions. On average, SDNE outperforms the other methods by a big margin. In conclusion, exploring deep learning based graph embedding methods is a valuable direction for future work.

\subsection{Sensitivity to Graph Size}
Real graphs vary in the number of nodes and edges. On the one hand, representing larger graphs in low dimensions requires more compression. On the other hand, many data driven approaches require data to learn the parameters and thus may not perform well for smaller graphs as the data may not be sufficient to learn their parameters. To study the effects of graph size on the embedding performance, we evaluate the performance of methods on graphs of different sizes ranging from 256, 512, and 1024 (as shown in Figures~\ref{fig:ben_real} and \ref{fig:ben2_real}).

We observe that for most methods, the absolute $MAP$ and $P@100$ scores decrease as the size of the graph increases. For example, the average $MAP$ score of HOPE across all dimensions is 0.08, 0.06 and 0.04 for graphs of sizes 256, 512 and 1024, respectively. Similar trends are observed for Graph Factorization and Laplacian Eigenmaps. However, for SDNE, the performance decreases for 512 nodes and increases for 1024 nodes. The reason may be that the deep neural network has more parameters to store information about larger graphs and benefits from more data. Another key observation is that although the absolute $MAP$ values decrease, the performance relative to the random baseline improves. As the number of nodes increases, the link prediction task becomes inherently more difficult. With more nodes there are more candidate links to choose from, thus leading to a lower absolute performance. In general, the increase of performance relative to the random random baseline suggests that the state-of-the-art embedding approaches better capture the graph structure when there is more information about the links.

\subsection{Sensitivity to Graph Density}
A wide variety of network graphs in real world are sparse. However, the factor of sparsity varies depending on the domain of graphs and other connectivity patterns in the graphs. We now study the effect of graph density on the performance of graph embedding methods (Figures~\ref{fig:ben_real} and \ref{fig:ben2_real}). 

In Figure~\ref{fig:ben_real} we observe that overall the performance of embedding methods in terms of $MAP$ is not sensitive to graph density. $MAP$ values of Graph Factorization increases with increasing graph density but for other methods the trend is not significant. Further, with higher density the likelihood of correct prediction increases so the trend in Graph Factorization may not signify an increase in performance with increase in density. From Figure~\ref{fig:ben2_real} we observe that the models perform poorly on $P@100$ score for lower density. Thus, the methods make top link predictions with more accuracy when the graph is relatively denser.

\subsection{Sensitivity to Embedding Dimension}
An inherent parameter of every embedding approach is the number of dimensions in the embedded space. The embedding dimension controls the amount of overfitting incurred by the method. A low number of dimensions may not be sufficient to capture the information contained in the graph. A high number of dimensions may overfit the observed links and thus perform poorly on predicting the missing links. 

The optimal embedding dimension depends on three factors: (i) the input graph, (ii) the embedding method, and (iii) the evaluation metric. Theoretically, graphs with less structure and higher entropy require higher dimensions. Furthermore, embedding methods which are capable of capturing more complex information may require higher dimensions to store the information. Also, predicting top links requires a higher level understanding of the graph and thus requires lower dimensions. Finding the optimal embedding dimension for a given graph requires extensive experimentation.

Figure~\ref{fig:ben_real} shows the $MAP$ performance of the graph embedding methods when varying the embedding dimension from $2^4$ to $2^7$. These results provide us with interesting insights. The performance of SDNE improves proportionally when increasing the embedding dimension. Laplacian Eigenmaps performs best for lower dimensions and quickly overfits for higher dimensions. This is surprising as deep neural networks models are prone to overfitting. But as deep neural networks are capable of capturing more complex information, they benefit from higher dimensional representation. Laplacian Eigenmaps only captures the first order proximity and hence overfits for larger dimensions. This is not surprising as the embedding vector stores first order proximity with higher precision. HOPE and GF also improve their performance with increasing dimensions, but HOPE has higher gains comparatively.

Figure~\ref{fig:ben2_real} shows the $P@100$ performance. The SDNE upward $MAP$ trend is consistent across the metrics and we observe similar increase for $P@100$. However, we observe that the performance of HOPE deteriorates with increasing dimensions. Graph Factorization's performance is almost constant across all embedding dimensions. This follows the intuition that predicting top links does not require higher dimensions.

\section{Python Library for GEM-BEN}~\label{sec:library}
We have created a pip installable Python library for graph embedding method benchmark called gemben (https://pypi.org/project/gemben). The documentation of the gemben library can be found at https://gemben.readthedocs.io. A simple usage example for linux is as follows:
\begin{python}
$ virtualenv -p  python3 ./venv #create a virtual environment
$ source ./venv/bin/activate #activate the environment
(venv) $ pip install --upgrade pip
(venv) $ pip install gemben #install gemben
(venv) $ python3 #launch python
#import the experiment module
>>> from gemben.experiments.experiment import exp
#experiment on social domain with all baselines algorithms
>>> test = exp(domain="social", method="all")
#run the experiment
>>> test.run()

\end{python}

\section{Conclusion and Future Work}~\label{sec:conclusion}
This work introduced a benchmark for graph embedding techniques. \textcolor{black}{This benchmark rests on a corpus of 100 real-world graphs, which is largely a subset of the popular CommunityFitNet corpus~\cite{ghasemian:etal:2019} with additional graphs drawn from the Stanford Network Analysis Project (SNAP)~\cite{leskovec2016snap}, and provides a broad evaluation of state-of-the-art embedding approaches against traditional approaches.} We established that graph embedding approaches outperform traditional methods on a variety of different graphs. Further, we showed that the performance of the method depends on multiple attributes of the graph and evaluation: (i) size of the graph, (ii) graph density, (iii) embedding dimension, and (iv) evaluation metric. Further, the performance varies tremendously based on the domain of the graph. Finally, we presented an open-source Python library, named GEM-BEN, which provides the benchmarking tool to evaluate any new embedding approach against the existing methods.

There are three promising research directions: (i) automating hyperparameter selection, (ii) graph ensemble techniques, (iii) generating realistic synthetic data. During the experimentation, we observed that a considerable effort is spent in identifying optimal hyperparameters for graph embedding approaches. Automating this process can help in their evaluation. Secondly, we showed in this work that optimal approach for a graph depends on the domain and other properties. We can extend this concept to infer that different subgraphs of a graph may benefit from using different embedding approaches. Combining multiple approaches on a graph is non-trivial but can help improve link prediction performance. Finally, in the Appendix we show that graph embedding approaches do not perform well on synthetic graphs as the state-of-the-art synthetic graphs are too simple with few parameters. We believe generating more realistic synthetic graphs is an important problem which can benefit testing a new approach.

\section*{Acknowledgements}
This material is based upon work supported by the Defense Advanced Research Projects Agency (DARPA) under Agreement No. HR00111990017, and Contract No. W911NF-17-C-0094.

\section*{Appendix: Results on Synthetic Graphs}
\subsection{Synthetic Graphs}

Synthetic graphs are often used to simulate the properties of real networks. This approach is useful when analyzing network structure properties and network evolution process, especially when the real graph is very large or cannot be fully observed. Realistic synthetic graphs can provide reliable and practical statistical confidence on algorithmic analysis and method evaluation. Synthetic graphs are ubiquitous in various domains including sociology, biology, medical domain, internet, teamwork dynamics, and human collaborations.


Many graph embedding methods for link prediction, node classification, graph visualization, and graph reconstruction have been evaluated on synthetic graphs. For example, GEM~\cite{GOYAL2018} uses the Stochastic Block Model (SBM) graph for visualization and evaluation on link prediction, node classification, and graph reconstruction. Similarly, dyngraph2vec~\cite{goyal2019dyngraph2vec} also utilizes the SBM graph to evaluate different dynamic graph embedding methods on link prediction tasks. HOPE~\cite{Ou2016} uses synthetic graphs generated by the forest fire model to evaluate the embedding's ability to preserve graph high-order proximity.

Although synthetic graphs do not to always fully capture inherent properties of real graphs, they are still useful for analyzing the performance of graph embedding methods regarding different graph characteristics. In this appendix, we analyze the performance of traditional and state-of-the-art methods on synthetic graphs. Our goal is to establish a better understanding on how graph embedding methods perform on some of the most popular synthetic graphs available in the literature.

\subsection{Synthetic Graph Dataset}
We evaluate the graph embedding methods implemented in our benchmark on ten synthetic graphs. These synthetic graphs can be categorized in three domains: (i) social, (ii) biology, and (iii) internet. Social networks represent the relationships between users on online social platforms. They characterize friendship networks (e.g.,  Facebook), and follower networks (e.g., Twitter where links are explicit and direct between users). Social graphs follow small world properties and power-law distribution. Social networks usually contain hubs and community structures. It is well known that Random Geometric graph~\cite{penrose2003random}, Waxman graph~\cite{waxman1988routing} and Stochastic block model~\cite{Yuchung1987} can appropriately reveal such community properties typically found in social networks. Biology graphs can represent protein-protein interaction networks, gene co-expression networks, and metabolic networks. The Watts-Strogatz graph~\cite{watts1998collective} and Duplication Divergence graph~\cite{ispolatov2005duplication} are heavily used to simulate biological networks. Graphs in the internet domain are usually very large in size and scale-free such as the web graph in the World Wide Web. The Barabasi-Albert graph~\cite{barabasi1999emergence} and the Power-law Cluster graph~\cite{holme2002growing} have been used to simulate the power-law distribution found in internet graphs. In addition to the synthetic graphs described above which are designed to simulate certain type of real graphs, Leskovec et al.~\cite{leskovec2010kronecker} used the Stochastic Kronecker Graph to effectively model any real networks from different domains using four parameters through an iterative Kronecker product process.


We use the following  synthetic graphs to illustrate the inefficacy of existing data sets and to highlighting the performance gaps between different graph embedding methods:

\textbf{Barabasi-Albert graph}~\cite{barabasi1999emergence}: This method generates random graphs using a preferential attachment process. A graph of $n$ nodes is constructed by adding new nodes with $m$ edges which are connected to existing nodes based on their degree. The likelihood of connection to a node is proportional to its degree. The generated random graph has power-law properties similar to the ones found in real-world networks.

\textbf{Powerlaw Cluster graph}~\cite{holme2002growing}: It extends the Barabasi-Albert graph to include a triad formation step. When a new node $u$ is linked to an existing node $v$, with a probability $p$, $u$ is also connected to one of $v$'s neighbors . The likelihood of triad formation controls the clustering coefficient of the graph.

\textbf{Watts-Strogatz graph}~\cite{watts1998collective}: The model creates a ring graph of $n$ nodes. It then links each node to its $k$ nearest neighbors. Each link from a node $u$ is rewired randomly to a node $w$ with a probability $p$. The rewiring is done to create short paths between nodes. 


\textbf{Duplication Divergence graph}~\cite{ispolatov2005duplication}: This random graph model is based on the behavior of protein-protein interaction graphs. The model starts with an initial graph and follows two steps for evolution: duplication and divergence. In the duplication step, a uniformly randomly chosen target is duplicated, and then connected to each neighbor of the target node. In the divergence step, each edge from the duplicate is removed with a probability $1 - p$. The retention probability of an edge is $p$.



\textbf{Random Geometric Graph~\cite{penrose2003random}}: It is a spatial network that starts with an arbitrary distribution of $n$ nodes in a metric domain. Then it creates an edge between any pair of nodes if their spatial distance is under a certain threshold. This model simulates the community structure within human social networks.

\textbf{Waxman Graph~\cite{waxman1988routing}}: This model extends the Random Geometric graph by adding edges in a probabilistic process. First, it uniformly places $n$ nodes in a rectangular space. If the distance of two nodes is within the neighborhood radius $r$, then it adds an edge with a probability $p$. The Waxman graph also demonstrates the community structure within the network.

\textbf{Stochastic Block Model Graph ~\cite{Yuchung1987}}: This random graph generator splits $n$ nodes into $m$ communities of arbitrary size. For each pair of vertices that belong to the community $C_i$ and $C_j$ respectively, this model connects two nodes under probability $p_{ij}$. In order to preserve the community structure, the in-block connection probability is higher than the cross-block probability.


\textbf{R-Mat(Recursive Matrix) Graph~\cite{chakrabarti2004r}}: this model is able to simulate any unimodal or power-law graphs with a few parameters using the recursive matrix. It recursively subdivides the adjacency matrix into four equal-sized partitions and distributes edges within these partitions with unequal probabilities.

\textbf{Random Hyperbolic Graph ~\cite{krioukov2010hyperbolic}}: this generation method builds a graph based on hyperbolic geometry. First, it randomly places nodes in a hyperbolic disk of radius $R$, and then connects each pair of vertices with an edge if their distance is less than $R$.

\textbf{Stochastic Kronecker Graph ~\cite{leskovec2010kronecker}}: similar to the R-Mat graph's recursive generation process, this model builds the graph's adjacent matrix from a $2 \times 2$ parameter matrix via iterating Kronecker product. Each component in the matrix is a real number between 0 and 1. Stochastic Kronecker Graph can simulate realistic graphs while preserving all common realistic graph properties.



\begin{table*}[]
\small
\begin{tabular}{|l|l|l|l|}
\hline
Domain                    & Domain Properties                              & Graph Generator              & Generator Properties                                                                                           \\ \hline
\multirow{4}{*}{Social}   & \multirow{4}{*}{community based graphs}        & Stochastic Block Model       & dense areas with sparse connections                                                                            \\ \cline{3-4} 
                          &                                                & Random Geometric Graph       & community structure                                                                                            \\ \cline{3-4} 
                          &                                                & Waxman Graph                 & community structure                                                                                       \\ \hline

 \multirow{3}{*}{Biology}  & \multirow{3}{*}{variant graph properties}                              & Watts Strogatz Graph         &                ring shape graph                                                                                                \\ \cline{3-4} 
                          &                                                & Duplication Divergence Graph &  simulate protein-protein interations via duplication and divergence                                                                                                   \\ \cline{3-4} 
                          &                                                & Hyperbolic Graph             & \begin{tabular}[c]{@{}l@{}}large networks, power-law degree distribution and high clustering\end{tabular}    \\ \hline

\multirow{3}{*}{Internet} & \multirow{3}{*}{power-law degree distribution} & Barabasi Albert Graph        & \begin{tabular}[c]{@{}l@{}}power-law degree distribution, small-world property\end{tabular}                  \\ \cline{3-4} 
                          &                                                & Powerlaw Cluster Graph       & \begin{tabular}[c]{@{}l@{}}power-law degree distribution, small-world property\end{tabular}                  \\ \cline{3-4} 
                          &                                                & R-Mat Graph                  & \begin{tabular}[c]{@{}l@{}}power-law degree distribution, small-world property, self-similarity\end{tabular} \\ \hline
                          
All                       &      simulate any realistic graphs                                          & Stochastic Kronecker Graph   &  build graph via iterating Kronecker product                                                                                                    \\ \hline

\end{tabular}
\end{table*}

\subsection{Results}

This section analyses the performance of different graph embedding methods and link prediction heuristic methods on link prediction task on eleven synthetic graph datasets described in section 6.2. We evaluate different link prediction approaches from various perspectives, including graph domain, graph size, node average degree and embedding dimension.


Figure 4 represents the link prediction performance $MAP$ scores of eight methods on all synthetic graphs from three different domains: social, biology and internet with varying graph size from 256 to 8192 nodes, varying graph embedding dimensions from 16 to 256, and varying average node degree from 4 to 12. While varying one parameter, we keep other parameters the same across different graphs. The $MAP$ score shown in the figure is averaged across all  synthetic graphs belonging to that domain. Figure 5 shows $P@100$ scores of all methods. Different from the apparent advantage of state-of-the-art graph embedding methods such as SDNE and HOPE on link prediction tasks on real graphs shown from Figure 2, we observe that classic link prediction heuristics such as Jaccard's Coefficient, Common Neighbors and Adamic-Adar can surpass some graph embedding methods on certain simple synthetic graphs dataset. Some synthetic graphs have naive structure and link prediction heuristics are capable of learning the network structure. Different graph embedding methods also perform differently in terms of domain, graph size, node degree and embedding dimension. 

Figure 5 shows concrete $MAP$ scores of those methods on three chosen synthetic graphs from three different domains, also with varying graph properties and embedding dimensions. The chosen graphs include Random Geometric Graph from social domain, Watts Strogatz Graph in biology and Powerlaw Cluster Graph belonging to internet domain. Figure 6 shows their $P@100$ performance. Seen from the figures, even with same graph sizes, node average degree and embedding dimension, those methods have diverse performance on link prediction. Some synthetic graph generators contain certain network structure characteristics, which enable heuristic based and graph embedding based methods to learn. However, some synthetic graphs such as synthetic internet graphs are generated in a more stochastic way, which are less structured and all methods fail to predict correctly.

\subsubsection{Domain Performance}
We divide all sythetic graph generators into three domains: social, biology and internet. Each domain contains three kinds of synthetic graphs. We add Stochastic Kronecker graph to each domain with specific parameters with respect to that domain.  

For social network synthetic graphs, heuristic method Jaccard's Coefficient and Common Neighbours perform the best on the link prediction task. As synthetic social graphs are usually community based graphs, including Stochastic Block Model, Random Geometric Graph, Waxman Graph, nodes within the same community are more densely connected than nodes outside. Jaccard's Coefficient and Common Neighbours are capable of capturing such inherent property that nodes with high degree are more likely connected with nodes with high degree within the same cluster. For graph embedding methods, Laplacian Eigenmaps has the similar MAP performance with best heuristics methods and HOPE is also good at predicting top 100 missing links. SDNE doesn't perform well on synthetic graphs probably due to that those graphs are simply structured and training links easily overfit the model.

When it comes to biology graphs, non-parametric methods Common Neighbor, Adamic-Adar and Jaccard Coefficients have consistently good performance. Biological graphs are usually various as objects and connections are different under different context. However, synthetic biological graphs are constructed in certain heuristic manner, which explains the good performance of the heuristic link prediction methods. Graph embedding methods are easily overfitted with such graphs. Some graph embedding methods such as HOPE can still overpass all methods at predicting top 100 nodes.  

For Internet graphs, Barabasi Albert Graph, Powerlaw Cluster Graph and R-mat Graph all have power-law degree distribution and small-world property. Heuristic methods Adamic-Adar has the best MAP score and Jaccord Coefficient achieves the best Precision@100 score. Like social and biological graphs, internet synthetic graphs are also built using such heuristics. Heuristic link prediction methods are suitable to explain edge generation rules. Graph embedding methods like Graph Factorization performs better on synthetic graphs in internet domain. 

\subsubsection{Sensitivity to Graph Size}
In our experiments, we test all methods on synthetic graphs from small size(1024 nodes), medium size(2048, 4096 nodes) to large size(8192 nodes). From three subplots in the first column in Figure 4 and 5, we notice that the absolute values of $MAP$ decrease when graph size increases accross all methods in three domains. However, some methods including Laplacian Eigenmaps and HOPE have higher $P@100$ scores when graph size grows. On one hand, the reason that $MAP$ score is lower on large graphs might be that with the size of graphs increasing, it is harder to predict the possible edges over more candidate edges. For graph embedding methods, embedding using the same dimension on larger-size graphs require more information compression and it results in relatively poor performance. On the other hand, the density of graphs increases while the graphs nodes increase and average node degree stay the same. Higher density provides more information for individual nodes and it helps to increase $P@100$ of link prediction.

\subsubsection{Sensitivity to Average Node Degree}
Instead of using graph density which is less controllable in the graph generators, we look into sensitivity of link prediction methods and graph embedding methods to average node degree of graphs. In Figure 4, we observe that most of methods get better MAP performance along with the increase of average node degree on social and biological synthetic graphs. When graph size is the same, the higher average node degree means each node has more edges. It is easier to predict possible hidden links. However, those methods' performance has opposite trend on internet graphs. Internet graphs might be more stochastic and noisy when the size becomes bigger. In figure 5, Hope and Laplacian Eigenmaps shows good performance on $P@100$. Graph embedding methods are good at predicting top links on denser graphs.

\subsubsection{Sensitivity to Embedding Dimension}
The traditional link prediction methods make predictions based on node similarity metrics and they are invariant to dimension. Graph embedding methods represent nodes in embedding space and different embedding dimension could generate embeddings with different quality. Seeing from Figure 4 and Figure 5, the higher dimension the graph embedding methods use, the better $MAP$ and $P@100$ scores they can get on the same graph. As all the graphs sizes are much larger than embedding dimensions, it requires less compression of graph information when embedding dimension is higher. Higher dimension preserves more features of graphs nodes and it results in better link prediction accuracy. For internet synthetic graphs, increasing embedding dimensions is a smaller boost and the reason might be that there are too much noise in the graphs.

\begin{figure*}
   \centering
  \includegraphics[width=0.95\textwidth]{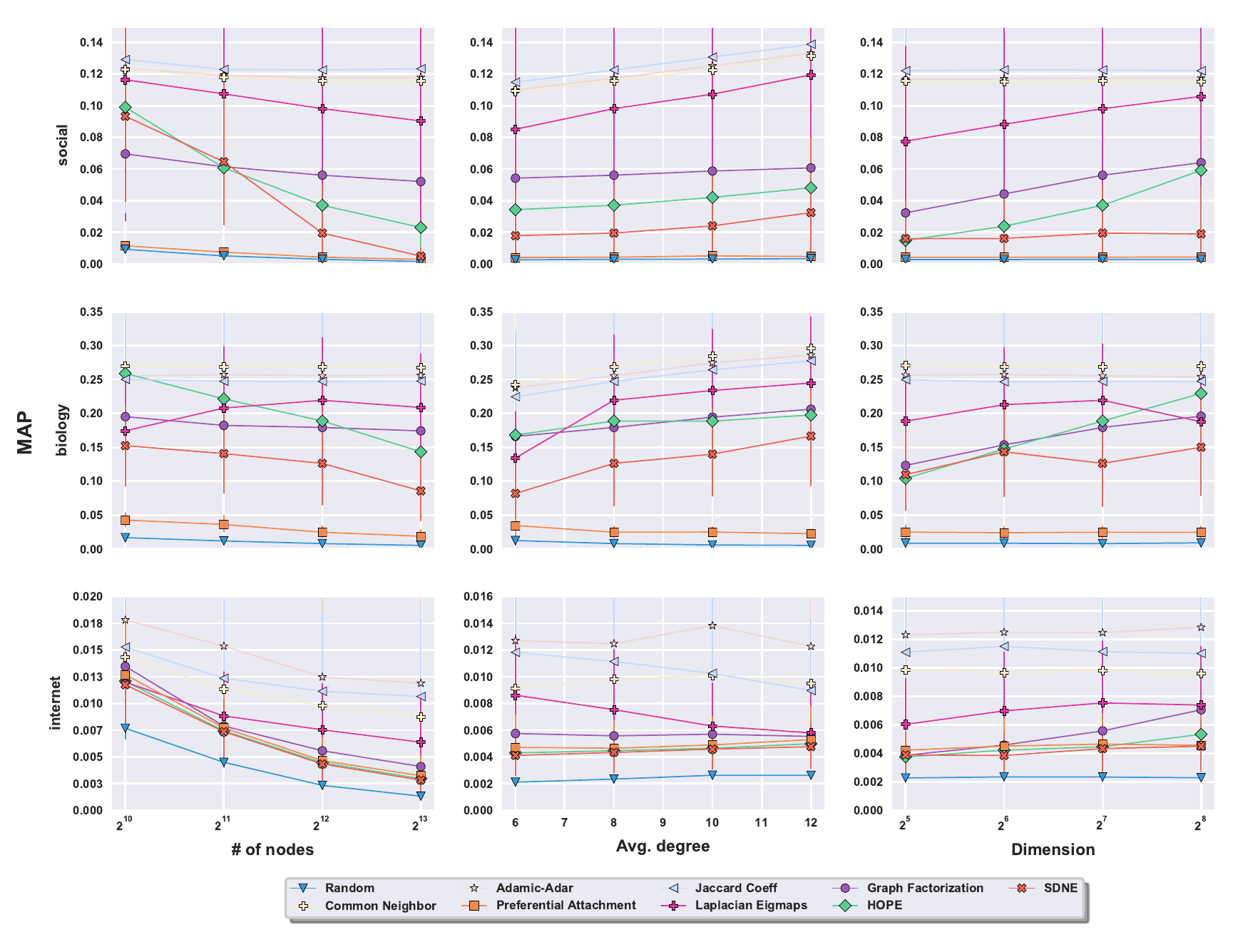}
  \caption{Benchmark Synthetic plot.}
  \label{fig:ben}
\end{figure*}

\begin{figure*}
   \centering
  \includegraphics[width=0.95\textwidth]{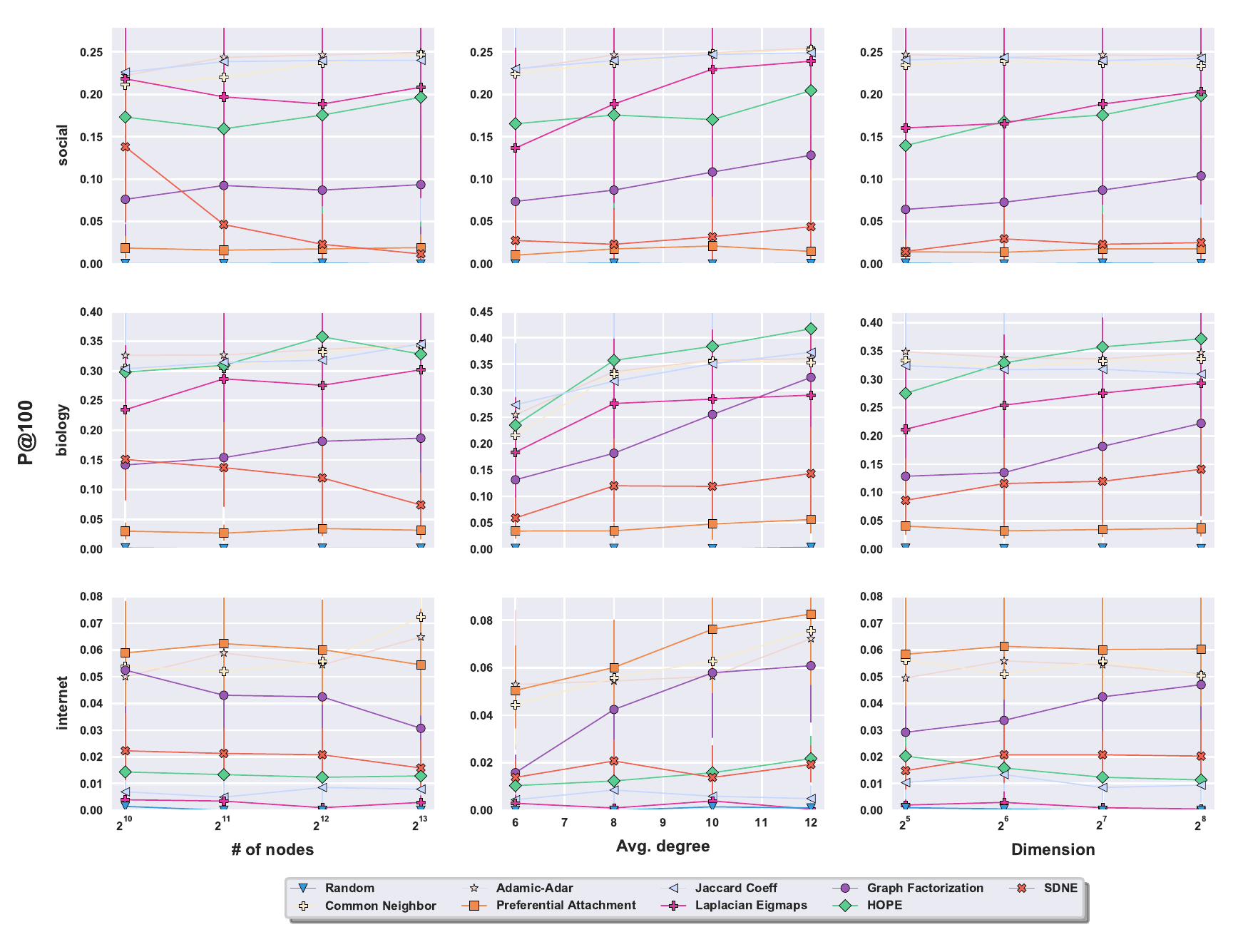}
  \caption{Benchmark Synthetic plot.}
  \label{fig:ben}
\end{figure*}

\begin{figure*}
   \centering
  \includegraphics[width=0.95\textwidth]{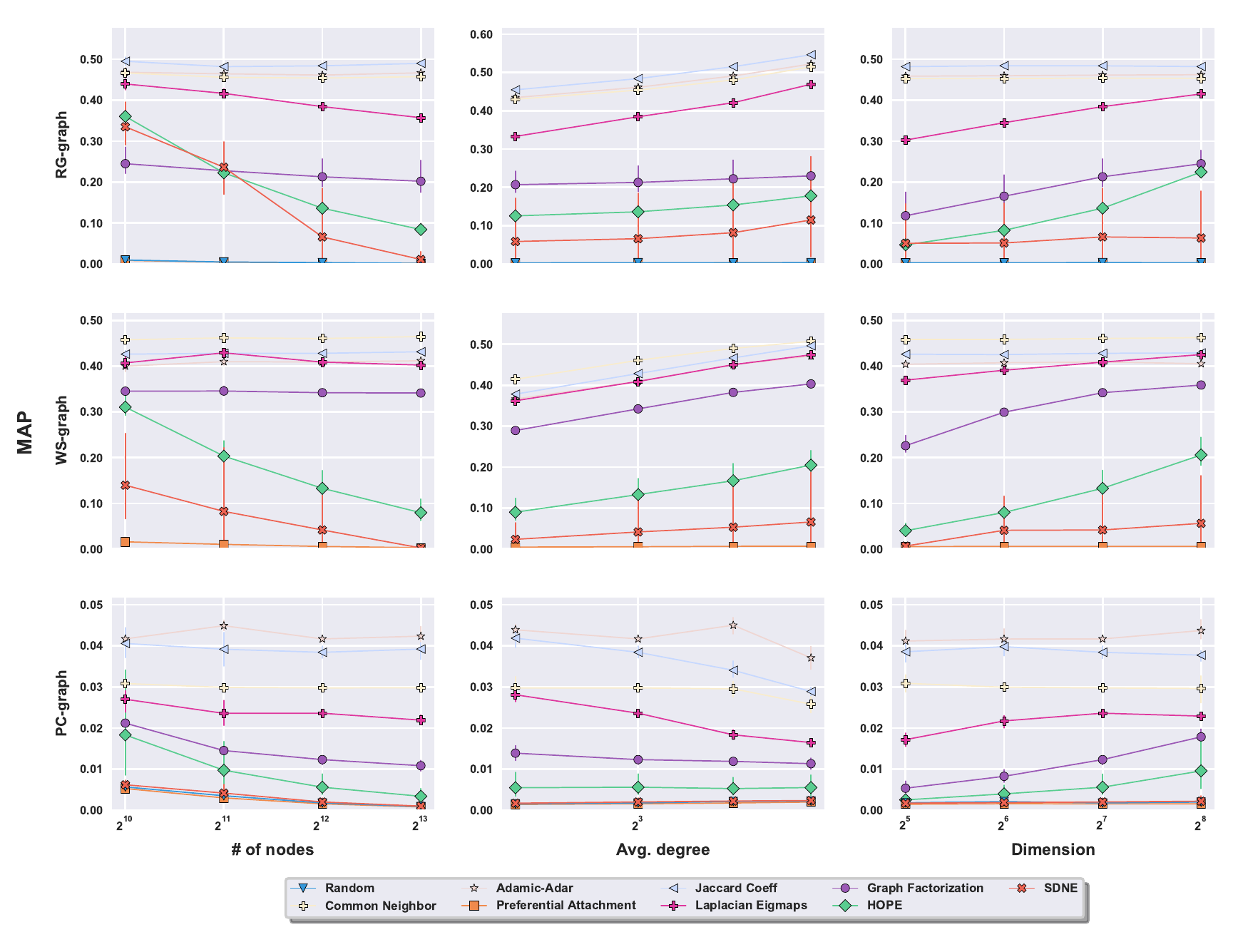}
  \caption{Benchmark plot for individual synthetic graphs.}
  \label{fig:ben_indi}
\end{figure*}

\begin{figure*}
   \centering
  \includegraphics[width=0.95\textwidth]{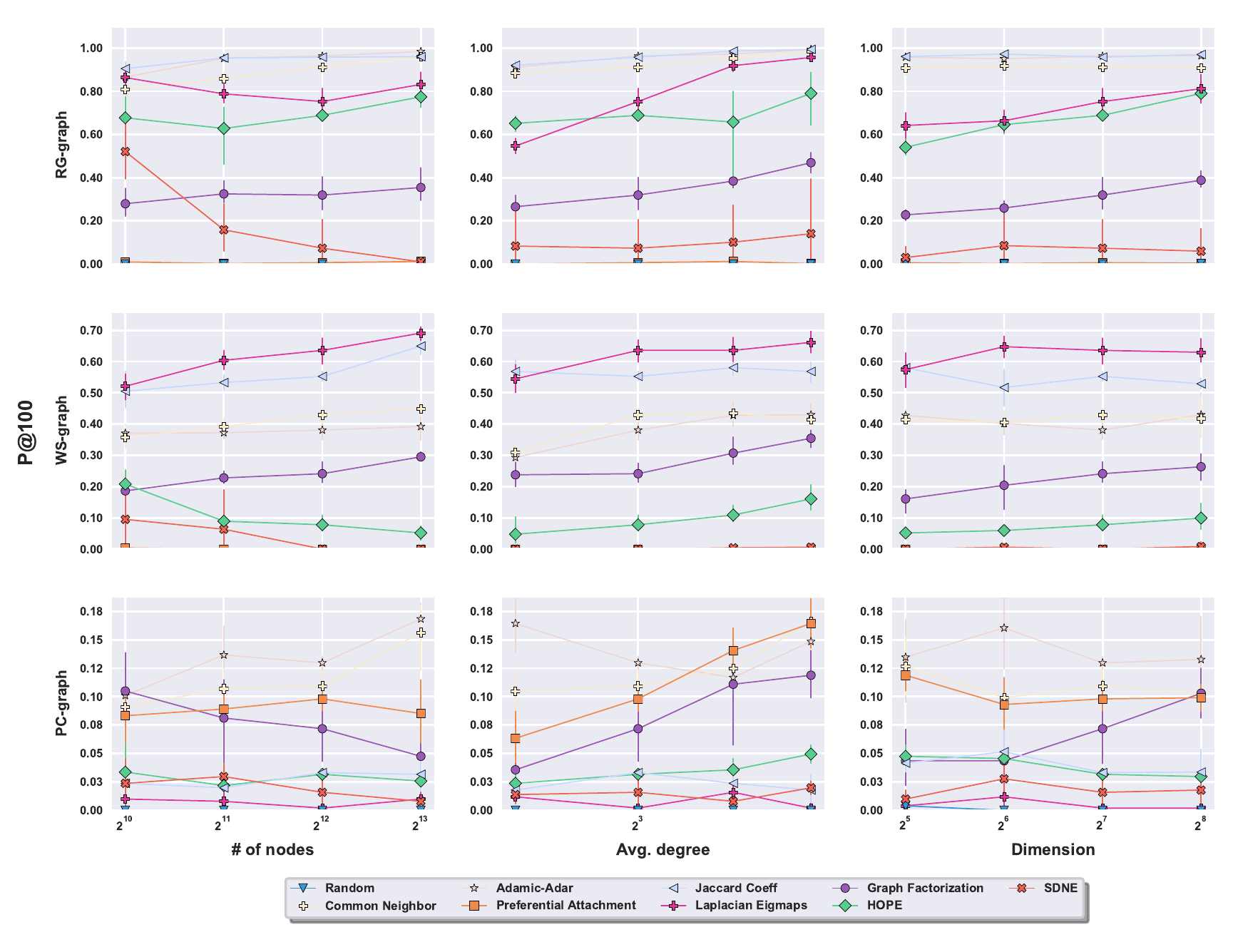}
  \caption{Benchmark plot for individual synthetic graphs.}
  \label{fig:ben_indi}
\end{figure*}
\balance
\bibliography{mybibfile}

\end{document}